# Flexible perovskite/Cu(In,Ga)$Se_2$ monolithic tandem solar cells

*Fan Fu,[1,2] * Shiro Nishiwaki,[1] Jérémie Werner,[2] Thomas Feurer,[1] Stefano Pisoni,[1] Quentin Jeangros,[2] Stephan Buecheler,[1] Christophe Ballif,[2,3] and Ayodhya N. Tiwari[1]*

[1] Laboratory of Thin Films and Photovoltaics, Empa-Swiss Federal Laboratories for Materials Science and Technology, Überlandstrasse 129, 8600 Dübendorf, Switzerland.

[2] Ecole Polytechnique Fédérale de Lausanne (EPFL), Institute of Microengineering (IMT), Photovoltaics and Thin-Film Electronics Laboratory, Rue de la Maladière 71b, 2002 Neuchâtel, Switzerland.

[3] CSEM, PV-Center, Jaquet-Droz 1, 2002 Neuchâtel, Switzerland

Corresponding author: fan.fu@empa.ch

We report a proof-of-concept two-terminal perovskite/Cu(In, Ga)$Se_2$ (CIGS) monolithic thin-film tandem solar cell grown on ultra-thin (30-µm thick), light-weight, and flexible polyimide foil with a steady-state power conversion efficiency of 13.2% and a high open-circuit voltage over 1.75 V under standard test condition.

## TOC GRAPHICS

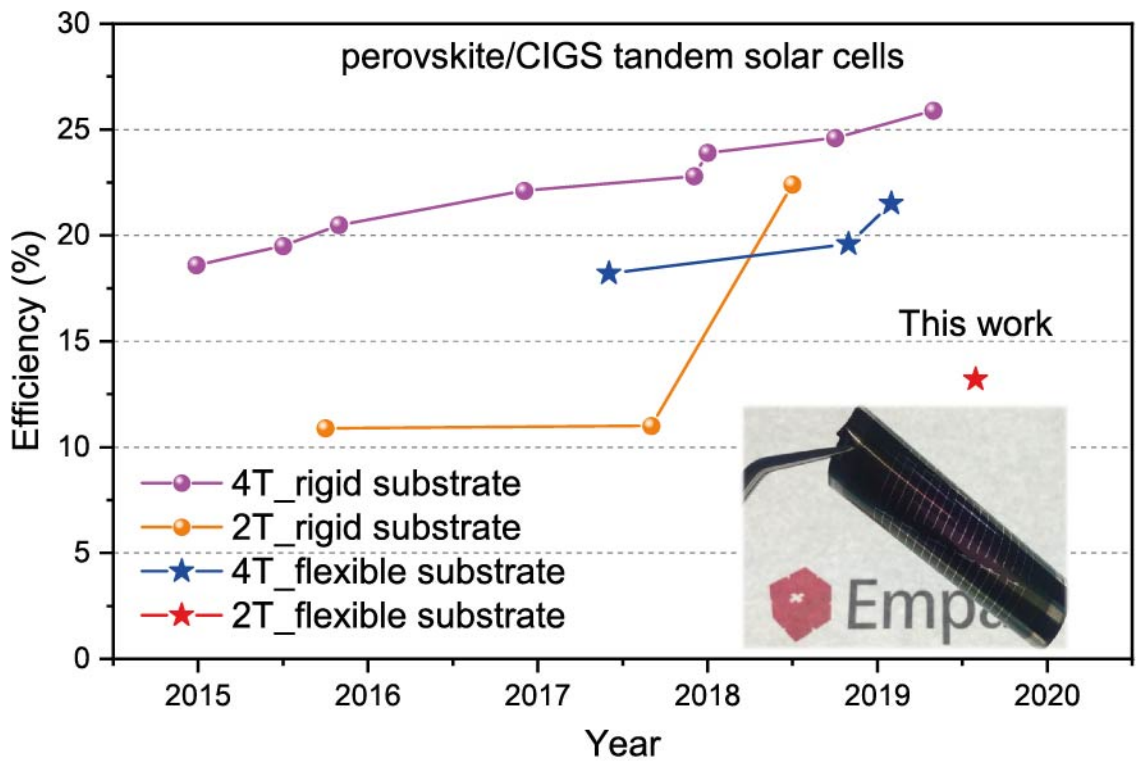

Roll-to-roll manufacturing of solar cells on flexible substrates is highly attractive for low-cost and high-throughput photovoltaic module mass production[1]. One of the most promising thin film solar cells is based on polycrystalline $Cu(In,Ga)Se_2$ (CIGS), demonstrating certified power conversion efficiencies of 20.8% on flexible polyimide foils[2]. Increasing efficiency at cell and module level is a straightforward and effective approach to reduce the levelized cost of electricity (LCOE) for CIGS photovoltaic technology. A promising way to bring the efficiency of CIGS solar cells to the next level is to construct tandem devices together with wide bandgap semiconductor absorbers. Attempts to fabricate tandem devices by combining CIGS bottom cells with various kinds of top cells, such as dye-sensitized solar cells (DSSC)[3,4], polymer solar cells[5], amorphous silicon a-Si:H[6], and CdTe solar cells[7], have yielded highest reported efficiency of 15%. The low tandem efficiency is primarily limited by the low efficiency and/or non-ideal bandgap of top cells. However, this situation has changed with the advent of perovskite solar cells[8,9], with certified power conversion efficiencies up to 24.2%[10]. Additionally, the wide and tunable bandgap over broad energy range makes perovskite ideal top cell candidates for tandem applications with narrow bandgaps bottom cells, such as crystalline Si[11,12], CIGS[13–15], Sn-based perovskite[16,17], and polymer[18] solar cells.

Among various kinds of tandems, perovskite/CIGS thin film tandems hold great promise as each single junction cell has reached >23% efficiency and the bandgaps of both absorbers can be easily tailored for current matching conditions[19,20]. Moreover, the operational stability of CIGS is way superior to narrow bandgap Sn-based perovskite solar cells[21–23], making it more suitable for highly efficient and long-term stable thin film tandem solar cells. Currently, the mechanically stacked perovskite/CIGS four-terminal tandems, where the perovskite top cells and CIGS bottom cells are fabricated individually and mechanically stacked together, have been intensively explored and already reached tandem efficiency up to 25.9%[13,24–28]. However, series connected 2-terminal tandem architecture, where top cell monolithically grows on top of the bottom cell, is seldom reported[15,29–31], although this tandem configuration is potentially preferred for commercialization in terms of reduced parasitic absorption losses and lower cost. This is mainly because of the large roughness of CIGS bottom cells, which makes it extremely difficult to deposit solution processed perovskite top cells on top of them without serious shunting problems. Very recently, this issue was tackled by Q. Han et al.[15] using mechanical polishing of the CIGS bottom cell to achieve smooth surface that allows subsequent perovskite processing, leading to a breakthrough of 22.4% monolithic perovskite/CIGS tandem solar cell with high open circuit voltage of 1.77 V. So far, all 2-terminal perovskite/CIGS tandems are deposited on rigid substrates. To exploit the full potential and advantages of perovskite/CIGS thin film tandems, it is highly desirable to develop perovskite/CIGS tandem solar cells on flexible substrates. This would open up the possibilities for high throughput roll-to-roll manufacturing and various attractive applications including the portable and wearable electronics, smart buildings, transportations, and aerospace, etc., where a combination of high efficiency, lightweight, and flexibility are important considerations.

In this work, we report a proof-of-concept two-terminal perovskite/CIGS monolithic thin-film tandem solar cells on ultra-thin (30 microns thick), lightweight and flexible polyimide foils, with a steady-state efficiency of 13.2% and a high open circuit voltage $V_{OC}$ of 1.751 V measured under standard test condition (25°C, AM1.5G, 100 mW $cm^{-2}$).

**Figure 1a** schematically illustrates the device architecture of the two-terminal flexible tandem solar cell. The perovskite top cell is monolithically grown on top of flexible CIGS bottom cell, and the top and bottom sub-cells are connected in series by recombination junction in between. The device structure of the flexible tandem employed in this study is as follows: polyimide substrate/Mo/CIGS/CdS/intrinsic ZnO (i-ZnO)/aluminum-doped ZnO (AZO)/poly(triaryl amine) (PTAA)/$CH_3NH_3PbI_3$/Phenyl-C61-butyric acid methyl ester (PCBM)/ZnO nanoparticles (ZnO-np)/AZO/Ni-Al metal grid/$MgF_2$. The fabrication of flexible CIGS bottom cell has been reported previously[2,32,33]. Due to the wavy polyimide substrate (**Figure S1**) and large surface roughness of CIGS bottom cells (**Figure S2**), it is quite challenging to deposit solution processed perovskite top cell directly onto flexible CIGS bottom cells without serious shunting problems. Although mechanical polishing has been employed to smoothen the surface of the CIGS bottom cell grown on a glass substrate, this method cannot be applied to flexible CIGS solar cell. Instead, we adopted the hybrid vapor/solution deposition approach, where a compact $PbI_2$ template layer was conformally deposited by thermal evaporation followed by spin coating of $CH_3NH_3I$ (MAI) solution, to enable the growth of conformal perovskite on top of flexible CIGS bottom cell. More detailed device processing parameters can be found in Methods. **Figure 1b** displays the photograph of flexible perovskite/CIGS monolithic tandem solar cells grown on a 5 cm × 5 cm polyimide foil. The cross-sectional scanning electron microscopy (SEM) image of the flexible tandem device is shown in **Figure 1c**, where the perovskite top sub-cell is conformally coated on top of the CIGS bottom sub-cell. As seen from the zoom-in area shown in **Figure 1d**, the PTAA thin layer is also conformally coated on top of the rough i-ZnO/AZO. **Figure 1e** presents the elemental depth profiling of the flexible tandem device by the time of flight secondary ion mass spectrometry (ToF-SIMS), which further confirms the designed layer structure.

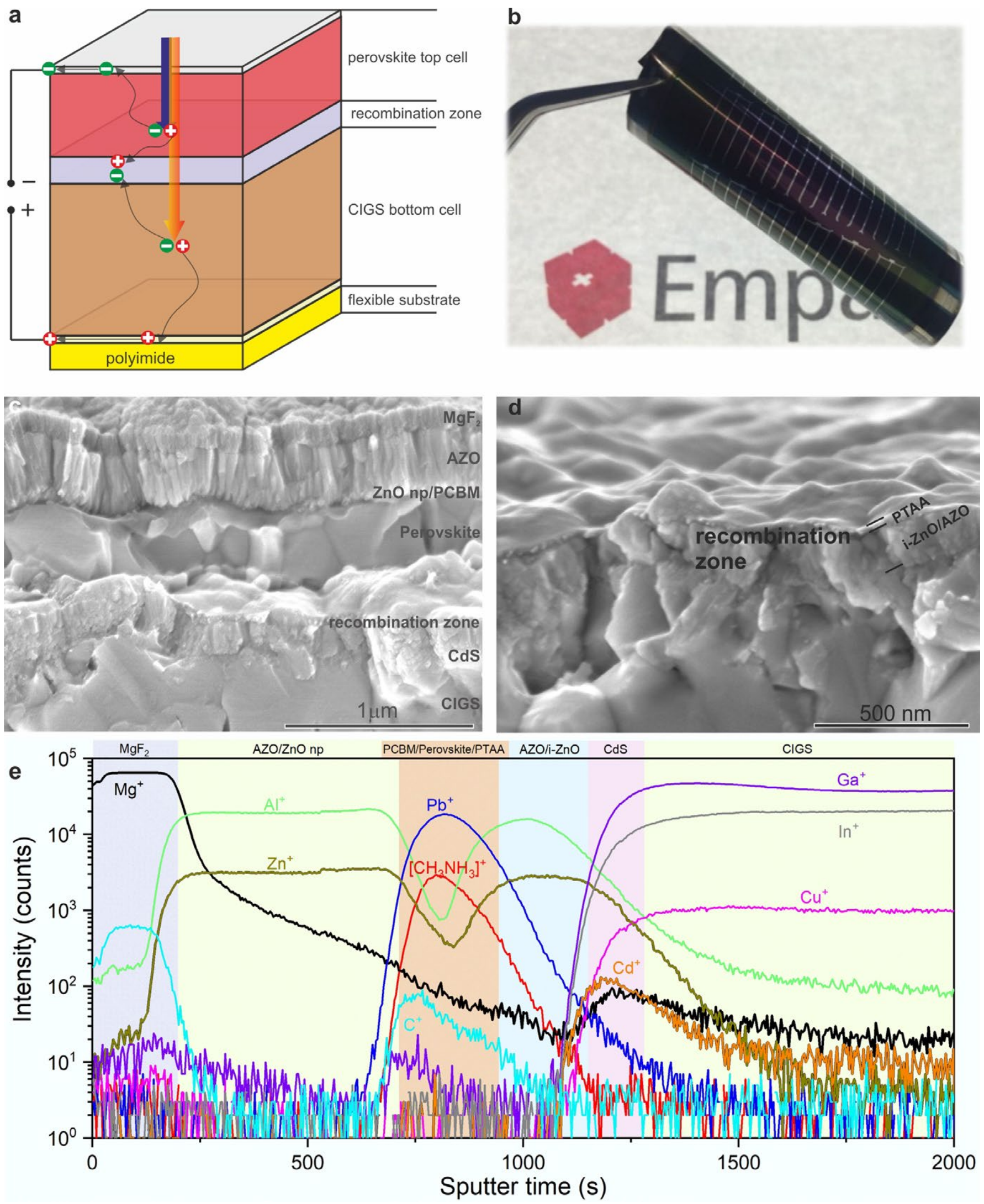

*Figure 1: a) Schematic illustration of the flexible 2-terminal perovskite/CIGS monolithic tandem device; b) Photograph of the flexible tandems grown on 5 cm × 5 cm polyimide foil; c) Cross-sectional scanning electron microscopy image of the flexible tandem solar cell; d) SEM image highlight the recombination zone (i-ZnO/AZO/PTAA); e) ToF-SIMS depth profiling of the flexible tandem cell.*

**Figure 2a** shows the photovoltaic performance of a small area (0.201 $cm^2$) flexible 2-terminal perovskite/CIGS monolithic tandem solar cell measured under standard test conditions (STC: 25 °C, simulated AM1.5G, 100 mW $cm^{-2}$). The flexible tandem device shows a $V_{OC}$ of 1.751 V, a short circuit current density ($J_{sc}$) of 16.3 mA $cm^{-2}$ and a fill factor (FF) of 46.4%, resulting in a power conversion efficiency ($\eta$) of 13.2% in the forward scan. In the reverse scan, $V_{OC}$ of 1.724 V, $J_{SC}$ of 16.4 mA $cm^{-2}$, FF of 42.3%, and $\eta$ of 12% are obtained. Although there is a small degree of *J-V* hysteresis, the cell measured at a fixed voltage of1.35 V (inset of **Figure 2a**) delivers a steady-state power conversion efficiency of 13.2% and current density of 9.8 mA $cm^{-2}$ for over 20 min of continuous illumination in ambient air. The steady state efficiency measurement is consistent with the reverse *J-V* scan. The external quantum efficiency (EQE) measurements of top and bottom cell in monolithic tandem configuration give an integrated $J_{sc}$ of 16 mA $cm^{-2}$ and 18.5 mA $cm^{-2}$, respectively. It is clear that the current density of the tandem device was strongly limited by perovskite top sub-cell.

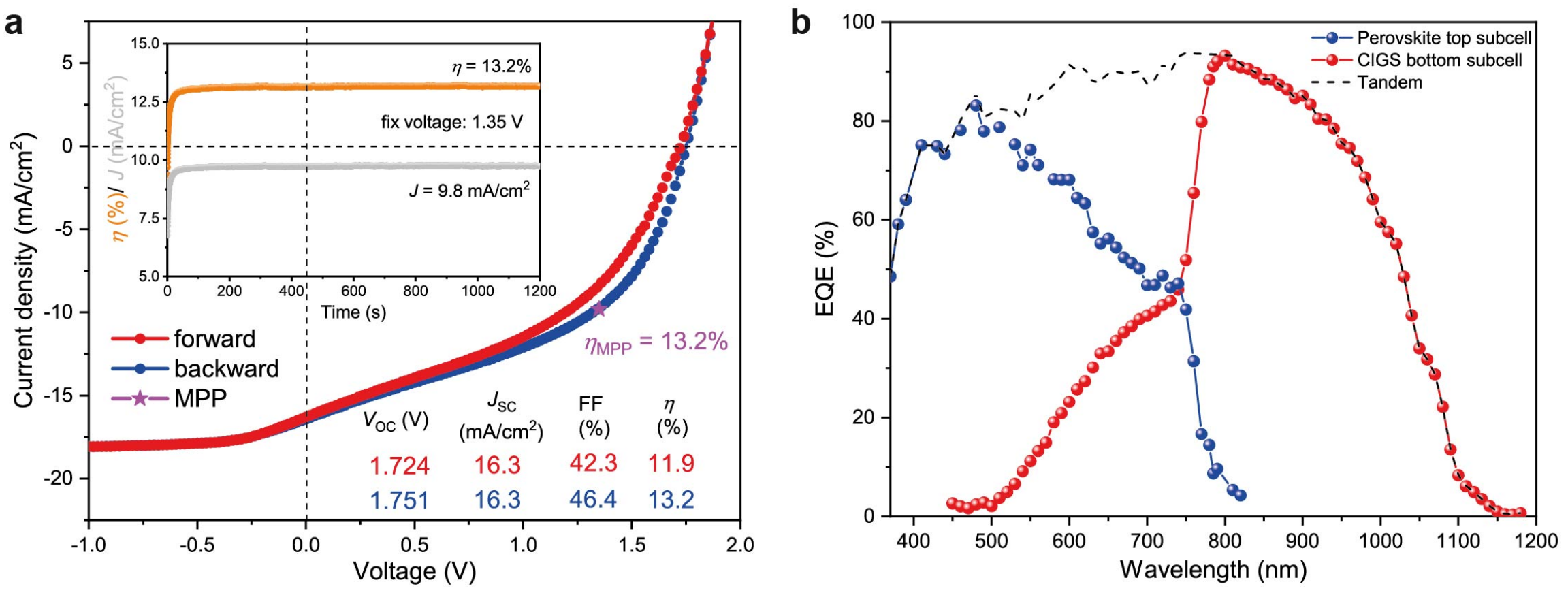

*Figure 2: a) J-V curves and steady-state power output (inset) of the 2-terminal flexile perovskite/CIGS monolithic tandem solar cell; b) Corresponding external quantum efficiency (EQE) spectra of the flexible tandem cell.*

The $V_{OC}$ of the flexible tandem cell is 1.751 V, which is almost the sum of $V_{OC}$ values from perovskite top cell (around 1.1 V)[14] and CIGS bottom cell (0.65 V from flexible CIGS after applying perovskite filter)[34], suggesting that i-ZnO/AZO/PTAA layers stack serves as an effective recombination junction. We note that the $V_{OC}$ achieved in the flexible perovskite/CIGS tandem cell is only 20 mV lower than that of the record perovskite/CIGS monolithic tandem on a rigid glass substrate. The $V_{OC}$ of the flexible perovskite/CIGS tandem cell is also comparable to the values reported for the state-of-the-art perovskite-based monolithic tandems as summarized in Table S1. However, the low $J_{SC}$ and FF strongly limit the power conversion efficiency of current flexible perovskite/CIGS tandem device to only 13.2%. The low $J_{SC}$ of the tandem cell is primarily due to the use of thin perovskite absorber (280 nm) layer. This leads to incomplete absorption of long wavelength photons between 500-750 nm as evidenced by the high EQE response of bottom CIGS shown in **Figure 2b**. Increasing the perovskite absorber thickness to around 500 nm is expected to enhance the $J_{SC}$ up to 19.1 mA/cm$^2$ as already demonstrated in our previous work[14]. In addition, the EQE of the perovskite top cell in the tandem configuration is still lower than that of perovskite single junction[14,35], suggesting a collection loss. This could be partially explained by the relatively poor morphology of perovskite absorber. As can be seen from the low-magnification cross-sectional SEM image shown in **Figure S3**, the grain size of perovskite is smaller than film thickness, resulting in plenty of grain boundary along the charge transport direction. The grain boundaries could act as non-radiative recombination centers, leading to reduced $J_{SC}$ and $V_{OC}$. The poor perovskite morphology is mainly ascribed to the short annealing time (1 min at 100 °C) of the perovskite absorber. The use of i-ZnO/AZO put a stringent constraint on the annealing temperature and annealing time of perovskite absorber. For example, the $MAPbI_3$ perovskite can easily decompose into hexagonal $PbI_2$ when annealing time is longer than 10 min at 100 °C (**Figure S4**). Our observation is consistent with previous results, where ZnO is known to degrade $MAPbI_3$ perovskite during thermal annealing.[36] Therefore it is highly desirable to employ thermally stable formamidinium/cesium mixed cation lead mixed-halide composition[37], and replace the AZO with ITO to increase the thermal budget allowed during annealing[29]. Although perovskite absorber can be conformally deposited via hybrid thermal evaporation/spin coating method, the spin coated PTAA, PCBM, and ZnO nanopartilces thin layers could introduce additional shunts in perovskite top cells. Efforts should also be devoted to substitute the spin coated electron/hole selective layers by vacuum-deposited counterparts without sacrificing the device performance.

To estimate the efficiency potential and to guide the rational design of efficient flexible perovskite/CIGS monolithic tandem solar cells, we employ transfer matrix optical simulations to estimate the optimal optical bandgap and thickness of perovskite absorbers that yields the highest tandem device performances using experimentally relevant parameters. In light of the above discussion, we propose a perovskite top cell device structure (**Figure 3a**) that could be feasibly conformally deposited on top of flexible CIGS bottom cells with significantly improved thermal stability at 150 °C. The AZO was replaced with ITO and (FA,Cs)Pb(I,Br) based perovskite composition was employed to enable high-temperature annealing in order to achieve the optimal morphology. The spin-coated charge selective layers (PTAA, PCBM, ZnO-np etc.) are replaced by vacuum deposited hole transporting layer NiO (magnetron sputtering) and electron transporting bi-layers C60 (thermal evaporation)/$SnO_2$ (atomic layer deposition) as reported previously[11,38]. Cross-sectional SEM images of the proposed tandem device structure are shown in **Figure S5**, where the perovskite absorber was conformally coated and exhibited a compact morphology after thermal annealing at 150 °C for 30 min in ambient air (relative humidity of around 40%). **Figure**

**3b** shows the optical simulations with different perovskite bandgaps. For each perovskite bandgap, the current matched $J_{SC}$ is optimized by varying the perovskite thickness. The ideal perovskite bandgap and corresponding thickness under current matched conditions are summarized in **Table S2**. For perovskite absorber with a bandgap of 1.6 eV, the current matched $J_{SC}$ is around 18.5 mA/cm$^2$ using 540 nm thick absorber layer. Taking the $V_{OC}$ of 1.751 V for flexible tandem cell already demonstrated here and a reasonable FF of 77.5%, it is quite realistic to achieve over 25% in flexible perovskite/CIGS monolithic tandem solar cells. The current matched $J_{SC}$ values maintain at 18.5 and 18.2 mA/cm$^2$ when increasing the perovskite bandgap to 1.66 eV and 1.73 eV, respectively. Assuming a $V_{OC}$ of 1.23 to 1.26 V for similar bandgaps reported in the literature[39,40], efficiencies up to 27% is practically achievable. Further efficiency enhancement should focus on tuning perovskite bandgap and reducing parasitic absorption loss. Thinning down the front TCO thickness and C60 thickness without sacrificing electrical properties could lead to higher current match $J_{SC}$ at around 19.5 mA/cm$^2$, which could yield 30% perovskite/CIGS flexible tandem efficiency in the future.

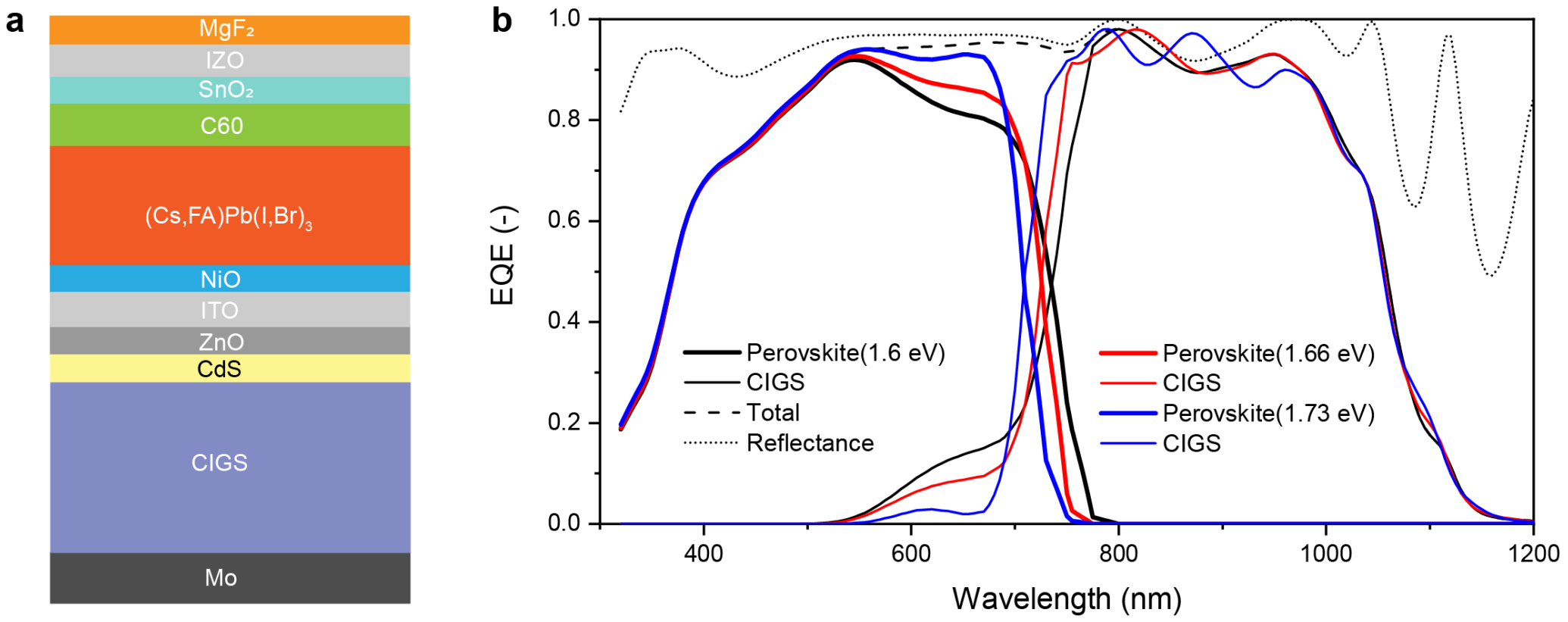

*Figure 3: a) Device structure used for optical simulation; b) Simulated EQE spectra of perovskite/CIGS monolithic tandem solar cells under current matching condition. Perovskite bandgap and corresponding thickness are varied to meet current matching critieria.*

In summary, we have demonstrated a proof-of-concept flexible perovskite/CIGS monolithic thin-film tandem solar cells with a steady-state efficiency of 13.2% and a high $V_{OC}$ of 1.751 V. We analyzed the current limitations and outline a viable pathway towards >25% efficient lightweight and flexible perovskite/CIGS monolithic tandem solar cells. Our results pave the way towards high throughput roll-to-roll manufacturing of high-efficiency, and low-cost thin-film tandem solar cells.

AUTHOR INFORMATION

Corresponding Author

*E-mail fan.fu@empa.ch.

ACKNOWLEDGMENT

This work was supported by funding from Swiss Federal Office of Energy (SFOE)-BFE (project No.: SI/501805-01) and Swiss National Science Foundation (SNF)-Bridge (project No.: 20B2-1_176552/1). We acknowledge the access to the Scanning Probe Microscopy User Lab at Empa for the (AFM) measurements.

# Supplementary Information

# Flexible Perovskite/CIGS Monolithic Tandem Solar Cells

*Fan Fu,[1] * Shiro Nishiwaki,[1] Jérémie Werner,[2] Thomas Feurer,[1] Stefano Pisoni,[1] Quentin Jeangros,[2] Stephan Buecheler,[1] Christophe Ballif,[2,3] and Ayodhya N. Tiwari[1]*

[1] Laboratory of Thin Films and Photovoltaics, Empa-Swiss Federal Laboratories for Materials Science and Technology, Überlandstrasse 129, 8600 Dübendorf, Switzerland.

[2] Ecole Polytechnique Fédérale de Lausanne (EPFL), Institute of Microengineering (IMT), Photovoltaics and Thin-Film Electronics Laboratory, Rue de la Maladière 71b, 2002 Neuchâtel, Switzerland.

[3] CSEM, PV-Center, Jaquet-Droz 1, 2002 Neuchâtel, Switzerland

Corresponding author: fan.fu@empa.ch

**Experimental methods**

Fabrication of flexible 2-terminal perovskite/CIGS monolithic tandem cell

The CIGS bottom sub-cell was prepared on commercially available flexible polyimide film on the basis of the method reported previously[1–3]. The CIGS layer was deposited on Mo coated (600 nm thick, DC sputtering) polyimide by multistage coevaporation process at around 450°C. In order to improve the performances of the CIGS sub-cell, NaF- and RbF-post deposition treatments were employed. The CIGS cell was completed with chemical bath deposition of CdS and then RF sputtering of ZnO and ZnO:Al layers.

The perovskite top sub-cell was directly grown on top of flexible CIGS bottom sub-cell in a p-i-n configuration. First, hole transport layer was deposited by spin coating 100 µl of PTAA solution (5 mg $ml^{-1}$ in toluene doped with 1 wt% F4-TCNQ) at 6,000 r.p.m. for 45 s, followed by thermal annealing at 105 °C for 10 min. Afterwards, a 140-nm-thick $PbI_2$ compact film was thermally evaporated on top of PTAA. The deposition rate was controlled within the range 1–1.5 Å $s^{-1}$, and the deposition pressure was in the range 3–6 × $10^{-8}$ mbar. Subsequently, samples were transferred into the glovebox for further processing. The perovskite layers were formed by spin coating of $CH_3NH_3I$ (MAI) solution. Specifically, 1 ml of MAI solution (45 mg $ml^{-1}$ in isopropanol) was first spread on the $PbI_2$ surface, then immediately start the rotation at 6,000 r.p.m. for 45 s. The as-deposited films were annealed at 100 °C for 1 min on a hotplate. For the electron transport layer, 100 µl PCBM solution (20 mg $ml^{-1}$ in chlorobenzene) was dynamically spin coated at 4,000 r.p.m. for 45 s, followed by 1 min annealing at 100 °C. After cooling down, 100 µl of undoped ZnO nanoparticles (2.5 wt% in isopropanol) was spin coated on top of the PCBM at 4,000 r.p.m. for 45 s to form bi-layer electron transport materials. The ZnO nanoparticles were dried at 100 °C for 1 min. Finally, the samples were finished with a AZO front contact by RF-magnetron sputtering and a Ni–Al (50 nm/4,000 nm) metallic grid by e-beam evaporation. 105 nm $MgF_2$ was applied as anti-reflection coating.

The current density–voltage (J-V) characteristics of the flexible tandem solar cells were measured under simulated AM1.5G illumination using a Keithley 2,400 source meter. The illumination intensity was calibrated to 1000 W $m^{-2}$ using a certified single crystalline silicon solar cell. The J–V measurement is performed in both forward (form −1 V to 1.8 V) and backward (from 1.8 V to −1 V) directions separately without any pretreatment (for example, light soaking, holding at forward bias for certain time and so on). The scan rate and delay time are 0.3 V $s^{-1}$ and 10 ms, respectively. The steady-state efficiency as a function of time was recorded by keeping the device at fixed voltage of 1.35 V for 20 min in ambient air. The external quantum efficiency of the devices were measured with a lock-in amplifier. The probing beam was generated by a chopped white source (900 W, halogen lamp, 260 Hz) and a dual grating monochromator. A certified single crystalline silicon solar cell was used as the reference cell. When measuring perovskite top cell, the tandem cell was light-biased by a white light equipped with a long-pass (>780 nm) filter; when measuring CIGS bottom cells, the tandem cell was light-biased by a blue light.

The cross-sectional scanning electron microscopy (SEM) images of the flexible tandem were investigated with a Hitachi S-4800 using 5 kV voltage and 10 mA current. Pt was coated on top of the sample to avoid charging effect. Time-of-flight secondary ion mass spectrometry (ToF-SIMS) measurements were performed on a ToF-SIMS.5 instrument from IONTOF, Germany, operated in the spectral mode using a 25 keV $Bi3^+$ primary ion beam with an ion current of 0.7 pA. For depth profiling, a 1 KeV $Cs^+$ sputter beam with a current of around 50 nA was used to remove the material layer-by-layer in interlaced mode from a raster area of 300 µm × 300 µm. The mass-spectrometry was performed on an area of 100 µm × 100 µm in the center of the sputter crater. The atomic force microscopy image was acquired using AFM Bruker ICON in PeakForce QNM mode (based on PeakForce Tapping technology).

Table S1 | Summary of photovoltaic parameters of the start-of-the-art perovskite based 2-terminal tandem solar cells.

| Institute | Tandem cell | $V_{OC}$ (V) | $J_{SC}$ (mA cm$^{-2}$) | FF (%) | $\eta$ (%) | $\eta_{MPP}$ (%) | *Area* (cm$^2$) |
|---|---|---|---|---|---|---|---|
| **Empa/EPFL (This work)** | **Perovskite/CIGS (flexible)** | **1.751** | **16.3** | **46.6** | **13.2** | **13.2** | **0.201** |
| IBM[4] | Perovskite/CIGS (rigid) | 1.45 | 12.7 | 56.6 | 10.9 | N/A | 0.4 |
| KIST[5] | Perovskite/CISe (rigid) | 1.346 | 12.9 | 63.5 | 11.0 | N/A | N/A |
| UW[6] | Perovskite/CISSe (rigid) | 1.4 | 14.5 | 42.1 | 8.5 | 8.6 | 0.11 |
| UCLA/UF[7] | Perovskite/CIGS (rigid) | 1.774 | 17.3 | 73.1 | 22.4 | 22.4 | 0.042 |
| HZB[8] | Perovskite/CIGS (rigid) | 1.59 | 18 | 75.7 | 21.6 | 21.6 | 0.778 |
| UCLA[9] | Perovskite/polymer (rigid) | 1.52 | 10.1 | 67 | 10.2 | N/A | 0.1 |
| HZB/EPFL[10] | Perovskite/SHJ | 1.759 | 14 | 77.3 | 19.1 | 18.1 | 0.16 |
| EPFL/CSEM[11] | Perovskite/SHJ | 1.692 | 15.8 | 79.9 | 21.4 | 21.2 | 0.17 |
| Stanford/AUS[12] | Perovskite/SHJ | 1.65 | 18.1 | 79 | 23.6 | 23.6 | 1 |
| EPFL/CSEM[13] | Perovskite/SHJ | 1.788 | 19.5 | 73.1 | 25.5 | 25.2 | 1.42 |
| HZB[14] | Perovskite/SHJ | 1.76 | 18.5 | 78.5 | 25.5 | 25.5 | 0.77 |
| Stanford/ ASU[15] | Perovskite/SHJ | 1.77 | 18.4 | 77 | 25 | N/A | 1 |
| U. Nebraska-Lincoln/ASU[16] | Perovskite/SHJ | 1.8 | 17.8 | 79.4 | 25.4 | N/A | 1.423 |
| NREL/U. Toledo[17] | Perovskite/perovskite (rigid) | 1.942 | 15 | 80.3 | 23.4 | 23.1 | 0.105 |
| U.Toledo/ NREL[18] | Perovskite/perovskite (rigid) | 1.922 | 14 | 78.1 | 21 | 20.7 | 0.105 |
| NREL[19] | Perovskite/perovskite (flexible) | 1.82 | 15.6 | 75 | 21.3 | 21.3 | 0.058 |

**Table S2 |** Transfer matrix optical modeling of a 2-terminal perovskite/CIGS monolithic tandem solar cell with various perovskite bandgap and corresponding optimized thickness.

| CIGS bandgap (eV) | $J_{SC}$ of CIGS (mA cm$^{-2}$) | Perovskite bandgap (eV) | Perovskite thickness (nm) | $J_{SC}$ of perovskite (mA cm$^{-2}$) |
|---|---|---|---|---|
| 1.15 | 18.46 | 1.6 | 540 | 18.6 |
| 1.15 | 18.54 | 1.66 | 750 | 18.52 |
| 1.15 | 18.9 | 1.73 | 1200 | 18.24 |

We assume that the IQE = 1 for all absorbers and all the interface are flat (no surface roughness). All layers are coherent except for CIGS which is considered incoherent. The device stack and layer thickness used for simulation are as follows: $MgF_2$ (100 nm)/ IZO (100 nm)/$SnO_2$ (10 nm)/ C60 (15 nm)/ perovskite/NiO (20 nm) / ITO (50 nm)/ ZnO (50 nm)/ CdS (40 nm)/ CIGS (2000 nm)/ $MoSe_2$ (10 nm)/ Mo (500 nm) /borosilicate (0.7 mm). The refractive indices of different materials used for transfer matrix simulation were taken from Roamin et al.[20] and Werner et al.[21] 's work.

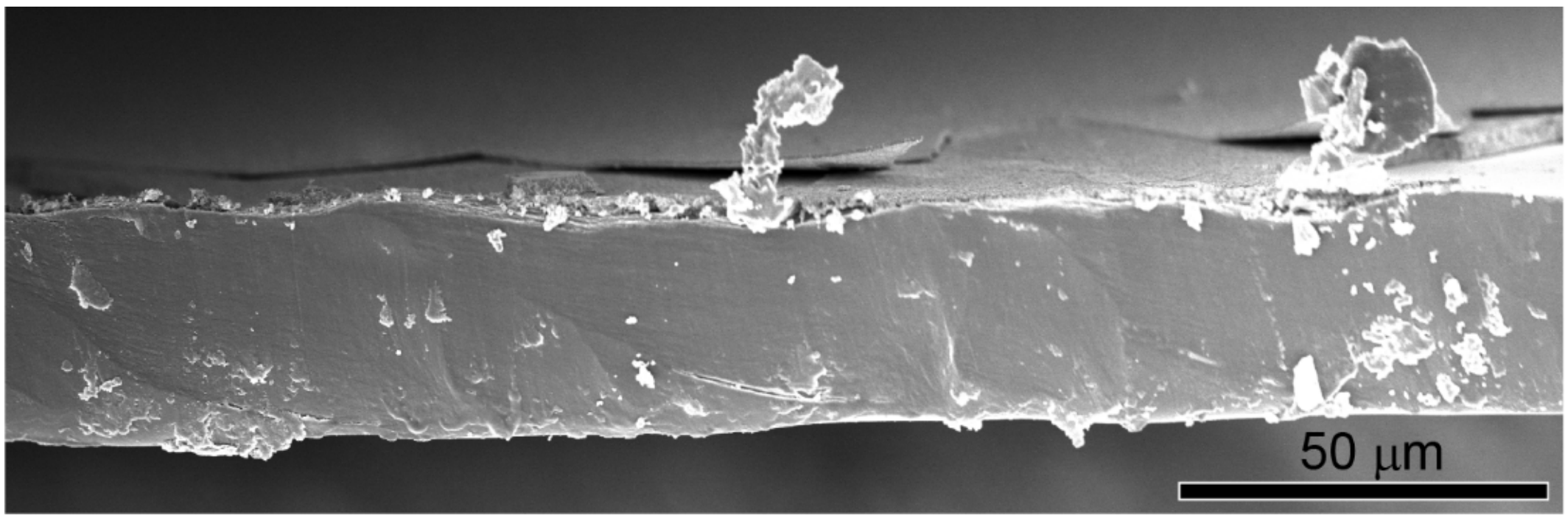

**Figure S1**. Cross-sectional SEM image of the 2-terminal perovskite/CIGS monolithic tandem solar cells fabricated on 30 μm thick flexible polyimide substrate. The polyimide substrate exhibit wavy morphology.

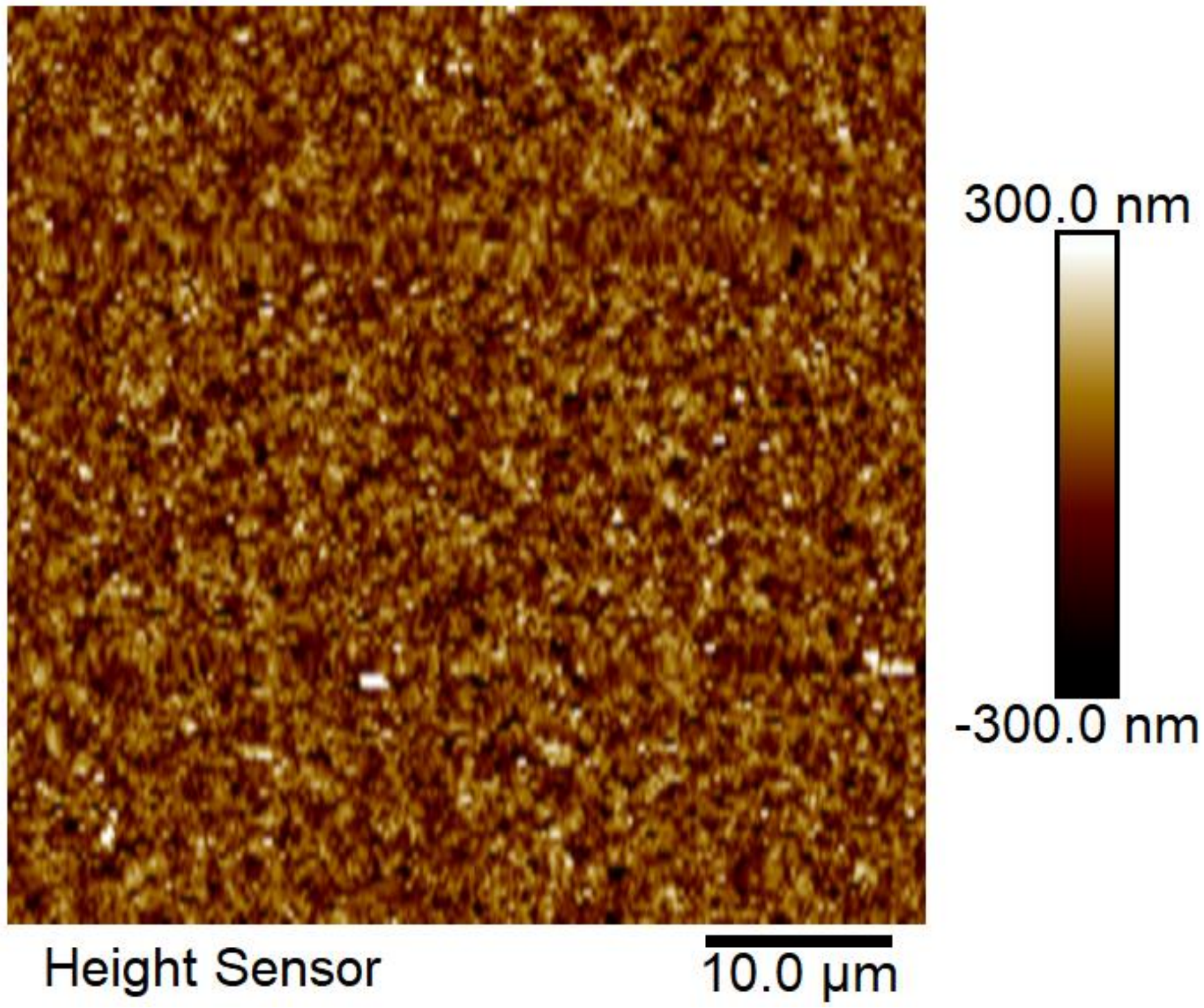

**Figure S2**. Atomic force microscopy (AFM) image of a typical CIGS absorber with root mean square roughness of 77 nm. The peak to valley distance could reach 1220 nm at some locations.

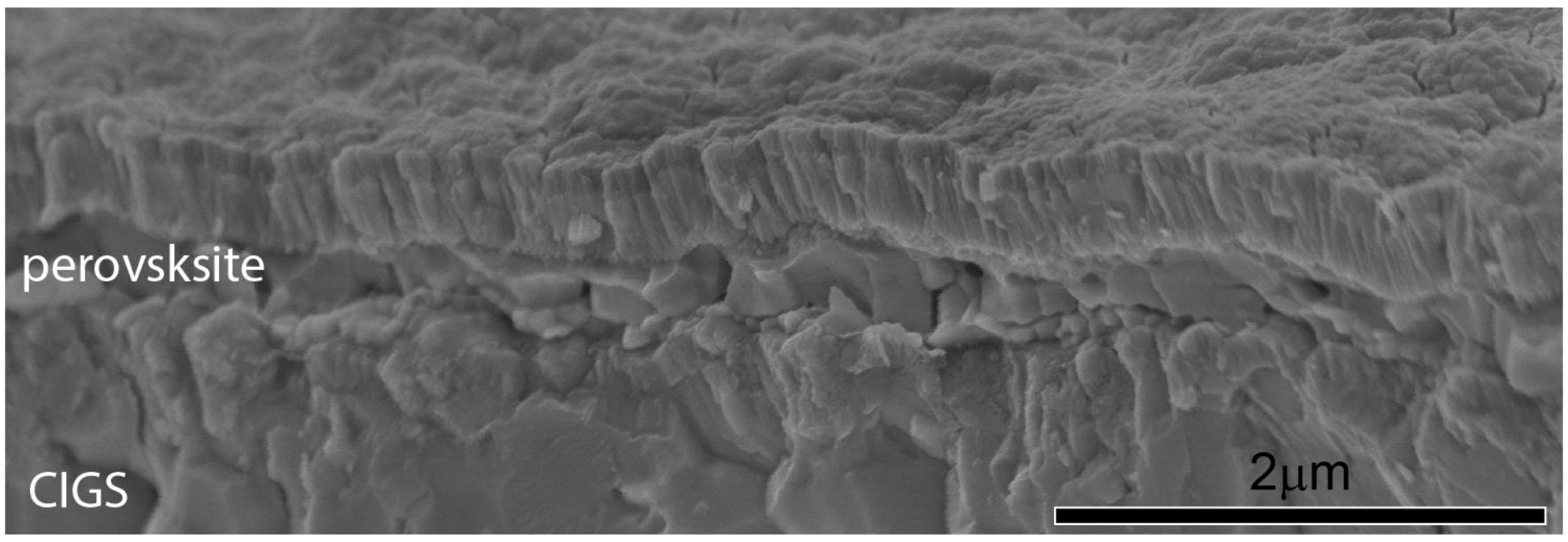

**Figure S3**. Cross-sectional SEM image of the 2-terminal flexile perovskite/CIGS monolithic tandem solar cell.

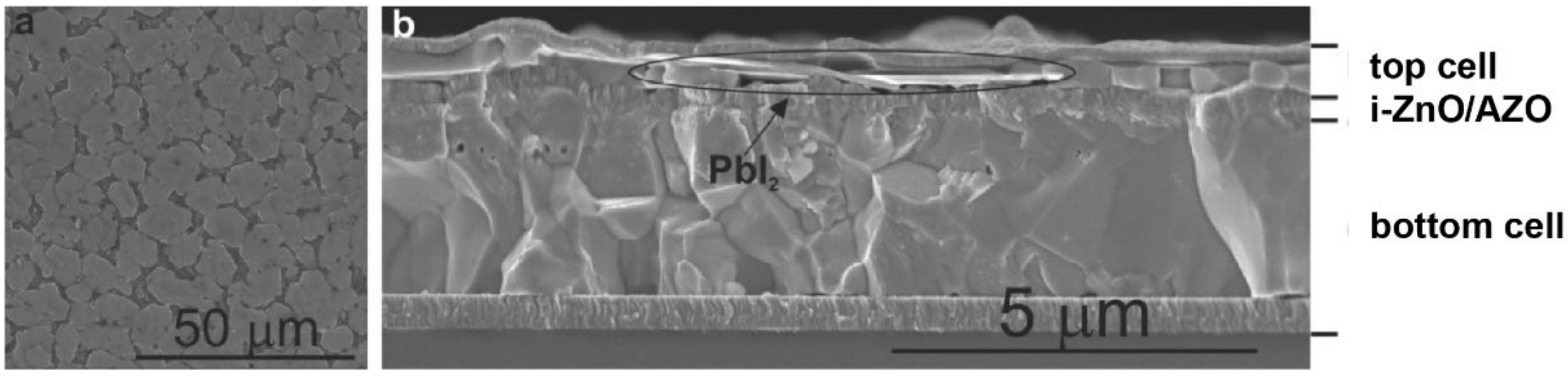

**Figure S4**. $MAPbI_3$ perovskite degraded on top of CIGS during thermal annealing. **a**, Morphology of perovskite after decomposition. The hexagonal shape indicates the formation of $PbI_2$. **b**, Cross-sectional SEM image of perovskite top cell degraded on top of CIGS. The perovskite absorber was annealed at 100 °C for 10 min.

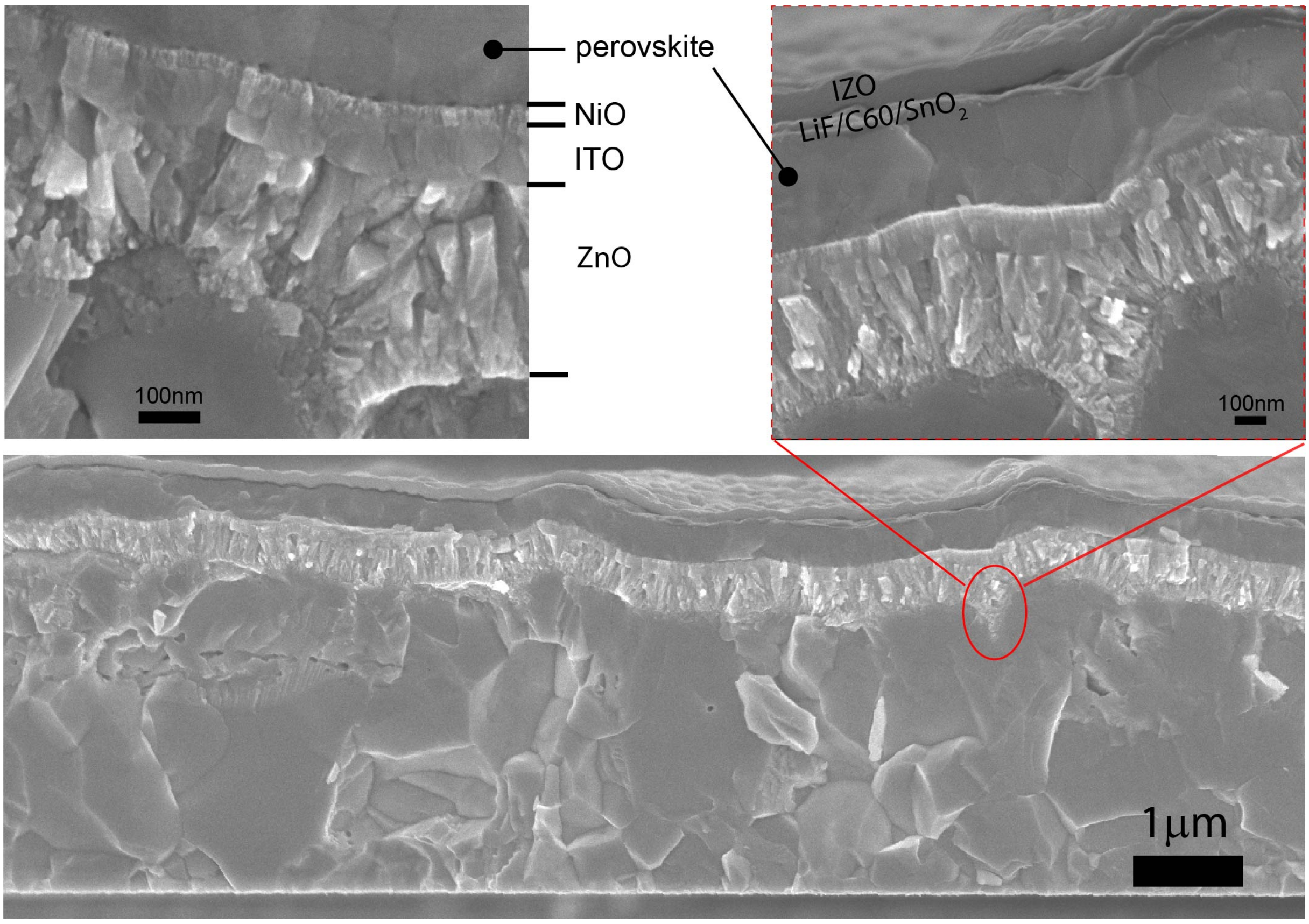

**Figure S5**. Cross-sectional SEM images of proposed 2-terminal perovskite/CIGS monolithic tandem solar cells. The perovskite absorber was prepared by a hybrid thermal evaporation/spin coating method[21] and demonstrated a compact morphology after 30 min thermal annealing at 150 °C in ambient air with 40% relative humidity.